\documentclass[a4paper,fleqn]{cas-sc}

\usepackage[authoryear]{natbib}
\usepackage{multirow}
\usepackage{xcolor}
\usepackage{array}
\usepackage{rotating}
\usepackage[dvipsnames]{xcolor}
\usepackage{float}
\usepackage[most]{tcolorbox}
\usepackage{cprotect}
\usepackage[shortlabels]{enumitem}
\usepackage[normalem]{ulem}

\newcolumntype{L}[1]{>{\raggedright\let\newline\\\arraybackslash\hspace{0pt}}m{#1}}
\newcolumntype{C}[1]{>{\centering\let\newline\\\arraybackslash\hspace{0pt}}m{#1}}
\newcolumntype{R}[1]{>{\raggedleft\let\newline\\\arraybackslash\hspace{0pt}}m{#1}}

\def\tsc#1{\csdef{#1}{\textsc{\lowercase{#1}}\xspace}}
\tsc{PB}

\begin{document}
\let\WriteBookmarks\relax
\def\floatpagepagefraction{1}
\def\textpagefraction{.001}
\shorttitle{New developments will change how we work fundamentally}
\shortauthors{P. Dueben et al.}

\title [mode = title]{Machine learning is revolutionizing weather forecasting – the next step is a change in how we work}                      

\author[1]{Peter Dueben}[type=editor,
                        auid=000,bioid=1,
                        orcid=0000-0002-4610-3326]
\cormark[1]
\fnmark[1]
\ead{peter.dueben@ecmwf.int}
\credit{Writing - Original draft preparation}
\affiliation[1]{organization={European Centre for Medium-Range Weather Forecasts},
                addressline={Robert-Schumann-Platz 3}, 
                postcode={53175}, 
                postcodesep={}, 
                city={Bonn},
                country={Germany}}

\author[2]{Peter Bauer}[type=editor,
                        auid=000,bioid=2,
                        orcid=0000-0002-3205-6055]
\fnmark[2]
\ead{bauerspeter@icloud.com}
\credit{Conceptualization of this study, Writing - Original draft preparation}
\affiliation[2]{organization={Max-Planck-Institute for Meteorology},
                addressline={Bundesstrasse 53}, 
                city={Hamburg},
                postcode={20146}, 
                country={Germany}}

\author[3,4]{Oliver Fuhrer}[type=editor,
                        auid=000,bioid=3,
                        orcid=0000-0002-0682-1374]
\fnmark[3]
\ead{oliver.fuhrer@meteoswiss.ch}
\credit{Writing - Original draft preparation}
\affiliation[3]{organization={Federal Office of Meteorology and Climatology MeteoSwiss},
                addressline={Operation Center 1, Airport}, 
                city={Z\"urich},
                postcode={8058}, 
                country={Switzerland}}
\affiliation[4]{organization={Institute for Atmospheric and Climate Science, ETH},
                addressline={R\"amistrasse 101}, 
                city={Z\"urich},
                postcode={8092}, 
                country={Switzerland}}

\author[5]{Nikolay Koldunov}[type=editor,
                        auid=000,bioid=4,
                        orcid=0000-0002-3365-8146]
\fnmark[4]
\ead{nikolay.koldunov@awi.de}
\credit{Writing - Original draft preparation}
\affiliation[5]{organization={Alfred-Wegener-Institute for Polar and Marine Research},
                addressline={Am Handelshafen 12}, 
                city={Bremerhaven},
                postcode={27570}, 
                country={Germany}}

\author[6]{J\o rn Kristiansen}[type=editor,
                        auid=000,bioid=5,
                        orcid=0000-0002-7210-9166]
\fnmark[5]
\ead{jorn.kristiansen@met.no}
\credit{Writing - Original draft preparation}
\affiliation[6]{organization={Norwegian Meteorological Institute},
                addressline={Postboks 43 Blindern}, 
                city={Oslo},
                postcode={0313}, 
                country={Norway}}

\cortext[cor1]{Corresponding author}

\nonumnote{This article compiles key areas where machine learning and recent technology developments will change methods by which weather prediction will be developed and implemented in the future.}
\nonumnote{The views expressed in this article are those of the authors and do not necessarily reflect the views, positions or policies of their affiliated institutions.}

\begin{abstract}
\mbox{}Following the success of machine learning in producing weather predictions with competitive skill compared to complex traditional systems, this article shifts attention from forecast output to the working practices that make prediction systems possible. We argue that machine learning and recent digital technologies will reshape the forecasting value chain: how models are coded and developed, how observations and Earth-system data are exploited, how data and computing are managed, how systems are verified, and how information is created, evaluated and turned into services. We discuss six non-exhaustive areas in which agentic software engineering, open and compressed data, shared verification workflows, interactive computing and generative methods may make modelling, evaluation and service creation faster, more interactive and more widely accessible. These changes will require weather and climate centres to adapt their infrastructures, data stewardship, trust and quality-assurance frameworks, skills and service delivery while maintaining scientific understanding, operational reliability, human expertise and their public-service role.
\end{abstract}


\begin{highlights}
\item Agentic coding, generative methods and modern software engineering will change the tools and workflows used to develop weather and climate models.
\item Open, compressed and better curated Earth-system data will change how observations are exploited, systems are verified and forecast information is accessed.
\item Weather and climate centres will need agile infrastructures, skills and trust frameworks while preserving reliability, human expertise and public-service delivery.
\end{highlights}


\maketitle

\section*{Introduction}\label{section:intro}
In the last three years, machine learning (ML) has revolutionised numerical weather prediction \citep{ZiedPangu,price2025probabilistic,kent2025skilful}. The initial push from technology companies has led to machine-learned forecast models running operationally as public services \citep{lang2026aifs}. Training efforts aside, ML produces often superior medium-range predictions at negligible time-critical production cost \citep{lam2023learning}. However, while the capability of end-to-end ML forecast systems is progressing \citep{alexe2024}, current systems still largely rely on initialisation and training data from traditional models, and ML predictions still produce a much smaller amount of output fields at coarser steps in space and time. The reliance on traditional models is crucial for driving science and understanding ML model performance, but this reliance ties the pace of ML model development to the rather incremental evolution of traditional systems. As the rate of score improvements from ML models for global NWP is starting to slow down after the initial surge, a dramatic, ML assisted change in the way we design cutting-edge systems to forecast weather and do climate projections is yet to come.

The revolution discussed here is not a single technical breakthrough, but a change in the operating model and forecasting value chain: who develops models, how data and compute are accessed, how systems are verified, how forecast information is turned into services, and how trust is maintained.
A big part of the transition concerns the way we work and how this is changing through ML and recent technological developments. The 'how' will determine the speed and cost-effectiveness of the adaptation to the ML revolution, as well as the extent to which digital technologies and software development practices from outside the weather domain can be adopted without introducing undesirable dependencies or loss of control over key capabilities and service quality. 

The traditional approach in Earth science has been that domain scientists with specialisations in mathematics, physics, chemistry and computer science jointly work on large mature codes. These are incrementally improved over time, tested using standard metrics, and then turned into data and products disseminated to users \citep{bauer2015quiet}. The transfer cycle from research to operations is typically around a year for weather and much longer for climate models. Today, this working model appears increasingly strained in a context where ML requires new skills across all science domains, software development is becoming faster-paced and aligned with industry standards, and large-scale computing architectures are driven more by AI than by traditional scientific requirements. As Moore-style cost-performance and Dennard-style power-density/clock-frequency scaling are expiring, and energy and data-movement costs of modern computing are exploding, it is evident that the next generation of supercomputers will not automatically provide more usable compute power for traditional physics-based weather and climate applications per Euro or Dollar spent \citep{shalf2020future}. The ‘how’ will therefore determine whether weather prediction and weather centres can remain productive and cost-effective.

We discuss six non-exhaustive areas, selected as a horizon scan because they cut across the forecasting value chain and are already visible in parts of current ML, software, data and computing practice. While the individual topics in isolation may not revolutionize weather prediction, their combination will fundamentally change the ways we work and how much impact is created. The discussion therefore combines developments that are already happening, emerging near-term trends, and longer-term scenarios. This is not meant as technological determinism \citep{smith1994does}, but as one plausible scenario for the future of our field: concrete enough to be useful for preparation, but not a forecast of an inevitable path. Most importantly, operational centres should start preparing for mixed human-AI system environments today because 'the how' will touch almost all areas of their work.

\section{Science \& codes: AI-assisted and agentic development workflows disrupt the working culture}\label{section:vibe}
Today, code development is performed manually by a large number of scientists and software developers affecting code quality, performance, portability across different high-performance computing (HPC) hardware configurations, and development pace. LLMs are already accepted as interactive research agents and text generators, but also emerge as primary code developers ('vibe coding', 'agentic coding') and integrators, code testers and workflow executors \citep{qian2024chatdev}. This trend has been seen in other domains earlier\cprotect\footnote{\verb|https://www.anthropic.com/research/vibe-physics|}, as their models and workflows are typically smaller when compared to today’s Earth-system models (ESM\footnote{The term Earth-system model is used here to denote the numerical model components representing the physical and chemical processes in atmosphere, ocean, land and cryosphere that are relevant for weather prediction from days to seasons and climate projections; it also includes the methodological framework to create the forecasts' initial conditions through the assimilation of observational data into models.}). But there are already examples showing that LLMs can understand the overall context of ESMs and data, and overcome our slow-paced manual approach \citep{pantiukhin2025accelerating,koldunov2026}.

To prepare large ESMs for the change, an important prerequisite is a focus on modularisation and complexity management to package smaller parts of the large code, and on clear Application Program Interfaces (API) communicating between the smaller packages and the larger code environment. These can then be independently compiled, run and optimised, continuously integrated (CI) and tested – also automatically using LLM generated workflows – to ensure code reproducibility, accuracy and stability. Current agentic\footnote{Throughout this paper, we use “agents” here for systems, often powered by Large Language Models (LLMs) or other foundation models, that can plan, call tools, act on code/data/workflows and iterate over multiple steps; this is distinct from LLMs as generative models alone.} systems often work through retrieval, search, tool-use and scoped context, not by loading entire codebases. Modularisation, clear APIs and searchable code structure make agentic development workflows more effective because relevant context can be retrieved, tested and modified without exposing the entire code base to the model at once. Modularisation will also enable the use of more diverse software solution options for model components. For example, there is no fundamental reason why an ESM should not use Python for one component and Fortran for another if this brings benefits in terms of performance or readability to the scientist who develops the specific piece of code. 

Many tasks that consume a significant amount of time for software engineers and scientists today are likely to be automated or semi-automated by AI agents that go through the full workflow of writing, running and testing independently. The fact that agents keep learning and adapting creates a feedback loop that is very powerful. The relevant loop is not single-shot code generation but a repeated process of planning, execution, evaluation/testing and adaptation, where compilers, type systems, tests, static analysis, benchmarks and CI provide feedback that agents can use to converge. Early software-agent benchmarks already evaluate language models on real repository issues and agent-computer interfaces for automated software engineering, showing both rapid progress and the present limits of these systems \citep{jimenez2024swebench,yang2024sweagent}. 

Coding with the help of LLMs already feels like working together in a mixed AI-human team. But this is only the beginning as agentic AI -- the step to task ML tools to autonomously solve problems -- will potentially again change how we create model code, develop workflows, build data pipelines, orchestrate compute, verify results and interact with users. It is even possible that AI agents will take over many science tasks in the prioritisation, selection and steering of scientific experiments. 

The effect will be that:
\begin{itemize}
    \item ESM development, testing and transfer across research, pre-operational and operational environments in mature agentic workflows can potentially become many times faster, because agents can iterate continuously across coding, testing, documentation, optimisation and porting tasks.
    \item From the human perspective, barriers between coding languages will likely become lower, because agents can translate, wrap, refactor and even choose implementation languages for individual tasks. Language and tooling choices will still matter operationally, but increasingly as inputs to agentic workflows: compilers, type systems, libraries, tests and CI provide the feedback and constraints that agents need for reliable correction and optimisation.
    \item Code adaptation to specialised hardware (and in particular GPUs) and for parallel processing, memory management and communication will be automated to the level where specific versions of the code can be generated for each particular HPC.
    \item AI agents will take over a significant amount of the workload of today's engineers and scientists, who will increasingly focus on managing, test development and monitoring, rather than coding and writing.
\end{itemize}

\section{Software: Industry-standard development and generic libraries}\label{section:library}
State-of-the-art physics-based ESMs are still only using links to software libraries that are either created in-house or target small-sized, very well defined mathematical tasks such as BLAS for basic linear algebra. More generic, external software dependencies are generally viewed as risky as their sustainability and fitness-for-purpose may not be guaranteed and because software dependencies are seen as strategic risk regarding lock-in effects. However, this paradigm is challenged as we move towards higher-level, abstraction-based programming, for example via domain-specific languages (DSL) such as GT4Py \citep{paredes2023gt4py}, and widely used programming languages such as Python and Julia \citep{wagner2025high}. Directly related to this is also the use of big software packages such as PyTorch and JAX \citep{fuchs2024torchclim,robinson2026advancing} that have been the basis of machine-learned weather models from the beginning of their developments.

The move towards industry standard software makes the use of ML tools for coding and agentic AI easier, as the tools have seen more context during training. For physical models, the role of specialized DSL-like abstractions may therefore evolve, with some of their current code-generation and portability functions potentially being replaced by agent-based workflows, while domain-level abstractions remain important as high-level interfaces for both humans and automated systems. Open-source frameworks, community-driven libraries and multi-platform deployment strategies may become increasingly important to preserve flexibility and independence while benefiting from rapid innovation in the broader AI ecosystem. The strategic issue is therefore not only whether external libraries are technically useful, but whether critical components remain inspectable, substitutable, portable and maintainable over operational timescales. Even PyTorch and JAX may not live up to this very high bar as their developments are driven by big-tech companies. 

Agentic software engineering changes what maintainability means: less of the burden may fall on individual humans being able to understand and edit every layer manually, while more of it falls on tests, specifications, provenance, interfaces and code structures that allow automated systems to inspect, modify, rewrite and verify changes reliably. Agents may also reveal defects in existing code bases as much as they introduce new ones, so the quality question shifts toward continuous automated testing, security scanning, provenance tracking and scientific validation. 

The offloading of tasks to external libraries or coding agents will make large ESMs easier to understand, modify and deploy, and thus more accessible. NWP centres will deploy more modern industry-standard software development strategies to focus their work more on the domain specific tools needed and to work with smaller development teams that are more independently targeting individual model components. This will allow for more continuous updating of model versions and the introduction of modern software development practices, focused code sprints, CI and the implementation of application-specific products. 

The effect will be that: 
\begin{itemize}
    \item Software developments can focus more on the scientific content allowing for a significant increase in productivity of staff.
    \item Development teams will stay small and focused, which will also improve productivity substantially.
    \item ESM development, deployment and testing can be shared more widely and become accessible to smaller and less specialized organisations, and countries.
\end{itemize}

\section{Data stewardship: Vast amounts of Earth system data discoverable, accessible \& open}\label{section:observe}
An important reason why the development of ML medium-range, global weather models was so fast and successful in recent years is the availability of open-source, easy-to-handle-and-access, high-quality data sets in form of reanalyses. Reanalyses use a single, stable system to assimilate observational data into models for several decades to produce an optimum estimate of weather evolution around the world based on the best available knowledge \citep{hersbach2020era5}. 

Increasingly, ML-methods that augment or even bypass the costly data assimilation step \citep{geer2021learning,aardvark,xu2025fuxi,fan2026physically} or directly produce analyses and forecasts from observations are being developed \citep{mcnally2024data}. Recent data-to-forecast systems also show progress towards cycling global short-range forecasting from raw satellite observations \citep{sun2024fuxi}. However, data assimilation is not only a computational step: it also provides observation quality control, bias correction, uncertainty estimation, observation impact assessment and dynamically balanced initial states. The dominant question is whether observations alone contain enough information to be competitive. If yes, the most important limitation is the access to as many observations as quickly as possible for both training and inference. While most of the traditionally assimilated data are available from open and free-access sources \citep{taalas2025free} today, many more need to be individually collected from sites with different access rights, procedures and latencies.

In the future, ML will be at the centre of observational data discovery, quality control and monitoring, and analytics, and ultimately the automation of these processes which may lead to a step-change for the use of non-traditional sources, commercial data, and internet-of-things (IoT) data. This will not always align cleanly with open-data principles: some high-value streams may remain proprietary, restricted by licences, or accessible only through commercial or national infrastructures. 

Some raw satellite datasets can be very large, for example from interferometers (data rates up to 2Gb/s\cprotect\footnote{\verb|https://user.eumetsat.int/resources/user-guides/mtg-irs-level-1-data-guide#ID-Dimensions|}) or synthetic-aperture radars (data rates up to 600Mb/s\cprotect\footnote{\verb|https://space.oscar.wmo.int/instruments/view/sar_c_sentinel_1|}), and therefore create transmission and analysis bottlenecks. However, in most cases ESM-simulated data volumes exceed those from observations. It will be of great importance to find joint solutions for edge computing (for high-data-rate instruments; \cite{leyva2023satellite}), streaming, post-processing and long-term storage that minimize data transfer. The standardisation of data curation protocols and formats (e.g. Zarr) will be key. Cloud-native scientific repositories and analysis-ready, cloud-optimised datasets provide one practical model for this shift \citep{abernathey2021cloudnative}.

Lossless compression methods find wide distribution in Earth-system modelling but only allow for limited compression ratios while lossy compression methods promise much higher gains but are less well accepted (except for fixed-point numbers in GRIB compression). This is because (i) the different physical fields of the models will need different loss definitions (e.g. using L1, L2, Linfinity norms) while the fields come with vastly different probability distributions (e.g. linear, Gaussians, log-normal, incl. zeros such as rainfall), (ii) diagnostics can be sensitive to small changes created by compression (e.g. long-term accumulated budgets when compression is causing a rounding-bias), (iii) many of the large datasets (in particular from reanalyses or climate simulations) will be used by many different applications in many different and unforeseeable ways, which compression loss may jeopardize. 

Widely used NWP data formats like GRIB and NetCDF already allow much lower floating-point data representations than 64 bits, and lossy file compression \citep{81320}. Many operational centres already use quantisation and packing in GRIB, including CCSDS packing, so compression is not a new concept in operational weather prediction. However, very sophisticated content-aware lossy compression methods are currently investigated by the community to expand this potential \citep{Huang2026}.
As an example, all data from the Copernicus Atmospheric Monitoring Service (CAMS) could be compressed by a factor of 17, while preserving 99\% of real information \citep{klower2021compressing}. This result is relative to a 64-bit baseline and dataset-specific; it should therefore be read as evidence of additional potential rather than as a universal operational storage multiplier. Recent work on highly compressed reanalysis datasets such as CRA5 similarly illustrates the potential for more aggressive, scientifically controlled compression \citep{han2025cra5}. ML is perfectly suited to extend and generalize this possibility because its main purpose is to predict data from learning its characteristics \citep{li2025lossless,yang2023introduction}. 

An equally important factor is an agile management of the location and access patterns of data so that not only data volumes but also data transmission rates are minimized \citep{stevens2024earth}. Data may also be streamed from live model simulations rather than stored.
Cost-effective, low-latency ML training requires new concepts for data distribution, as exercised by companies like Google or AWS. Multi-grid access to large datasets, for example for km-scale global model data, show that scientists can already access petabyte-datasets within hours as demonstrated during a recent World Climate Research Program hackathon\cprotect\footnote{\verb|https://www.dkrz.de/en/communication/news-archive/wcrp-global-km-scale-hackathon|} and the general developments for operational digital twins in DestinE \citep{doblas2026destination}. 

The primary objective is to create easy-to-handle interfaces for users and hide the data-handling overheads. This is already supported by community-wide open-data trends, including the WMO Unified Data Policy, ECMWF open data, and open-data/API initiatives at several national meteorological services such as MET Norway, MeteoSwiss and the UK Met Office\footnote{\textcolor{red}{\url{https://wmo.int/wmo-unified-data-policy-resolution-res1}; \url{https://api.met.no}; \url{https://www.meteoswiss.admin.ch/services-and-publications/service/open-data.html}; \url{https://www.metoffice.gov.uk/services/data/met-office-weather-datahub}}}. The Destination Earth (DestinE) \citep{wedi2023destination}, and Copernicus \citep{smart2017scalable} initiatives of the European Commission are already investing in this domain, as are the EVE \citep{stevens2024earth} initiative and commercial companies\cprotect\footnote{\verb|https://www.nvidia.com/en-us/high-performance-computing/earth-2/|}. Concrete technical examples include feature-based extraction systems such as Polytope \citep{leuridan2025polytope} and standards-based interfaces such as OGC API Environmental Data Retrieval and CoverageJSON \citep{ogc2024edr,ogc2023coveragejson}. To keep applications in step with data, users must be able to access all relevant datasets on leading HPC centres as well as public clouds with minimum effort. If a user is requesting a list of all different datasets that are available at a certain point in time and space, they should be able to get such a list within seconds and the requested data within minutes. 

It is also important to realise that many of the users might be AI agents rather than humans in the future. They need simple and clear APIs to data, tool discovery through Model Context Protocols (MCPs), and datasets enriched with the context that is needed to process diverse data effectively.

As data volumes grow and workflows shift toward remote access and streaming, the role of dedicated data stewards will become increasingly relevant or important. Rather than distributing large datasets to individual institutions, specialised centres could curate, maintain and serve high-value datasets through standardised interfaces. These data stewards would manage compression strategies, metadata, quality control and versioning while providing efficient access for models and diagnostics. This shift could reduce duplication, improve consistency and enable collaborative workflows across institutions. Access to curated datasets may therefore become more important than local data ownership and first commercial providers are already positioning themselves to fill the gap\cprotect\footnote{See for example \verb|https://www.earthmover.io/marketplace/|.}. 

The effect will be that: 
\begin{itemize}
    \item When using sophisticated lossy data compression, substantially more data may be stored, streamed and post-processed, with gains depending on the dataset, baseline precision, diagnostic requirements and application allowing data analysis to evolve at the same pace as model resolution, complexity, ensemble size and observational data proliferation.
    \item Data stewards will enable the community to access unified model/observation/(re-)analysis data handling service infrastructure as an open-source resource providing unified access to data and HPC.
    \item ML training can use all available data from both large, long-term repositories and distributed and diverse, and continually generated observation and simulation streams.
    \item An increasing share of data access will be by AI agents rather than direct human interaction which may also lead to increasing data volumes with a respective increase in cost for training.
    \item The huge financial and organisational investments in observational platforms (e.g. satellite programmes) produce a much higher return than today and allow exploitation of new data types from commercial providers, IoT data and large research experiments.
\end{itemize}

\section{Verification: Scientific components, user metrics \& benchmarks}\label{section:verify}
To build ML weather and climate models requires large high-quality training datasets that provide a “truth” for all fields that should be learned. The detailed evaluation of these models is essential for building trust with domain scientists and monitoring progress in predictive skill. This is similar to the long-standing community practices established by the WMO\cprotect\footnote{\verb|https://community.wmo.int/site/knowledge-hub/programmes-and-initiatives/wmo-integrated-processing-and-prediction-system-wipps/forecast-verifications|}. 

Since the emergence of ML in NWP, many training datasets and selected open diagnostic packages have been developed that are usable well beyond traditional NWP centres. This includes ML-skill benchmark tools and datasets such as WeatherBench \citep{rasp2023weatherbench,dueben2022challenges} and ClimateLearn \citep{NEURIPS2023_ed73c36e}. The ability to perform detailed diagnostics with these tools will improve further as datasets will grow and as observational datasets will become integrated (see Section~\ref{section:observe}) allowing for much more detailed model evaluation work across all applications. Comprehensive observation-based diagnostics have been a strength of data assimilation for decades \citep{cardinali2009monitoring,laloyaux2025using}, but the ease of use of such tools and datasets in the context of ML offers huge potential. 

ML model training, testing and tuning becomes much easier than possible for traditional numerical models because it evolves mostly around refining training datasets, ML model set-up including hyper-parameter selection and cost-function specification. The latter offers new possibilities as ML models can be tuned to targeted predictions. This is particularly relevant for local and sector-specific applications, such as flooding, wind energy yield, food production, and extremes \citep{asrade2026flood,camps2025artificial,abdelsattar2025evaluating,nosratabadi2021prediction}. A growing verification literature now tests whether these gains hold for extreme events specifically \citep{olivetti2024extremes,nath2026can}. Since verification is no longer only an external assessment step in AI systems, metrics, benchmarks and diagnostics shape the training process and therefore the model itself.
As it becomes easier to verify model dynamics for many different aspects and as workflows and trustworthy benchmarks can be automated, it is likely that AI agents will be able to test models more reliably and therefore work more independently.

An important negative aspect is that such verification focuses too much on the selected norms rather than physical realism and robustness, and that datasets are often not sufficient to generalise to all points in space and time also leading to worse predictions in areas with fewer observations. This contrast should not imply that physics-based models get a free pass: they also contain tunable parameterisations, systematic biases and score-oriented calibration choices. Both ML and physics-based models therefore need multi-faceted diagnostics for physical consistency, conservation, calibration, uncertainty, robustness, extremes and user-level outcomes.

It is therefore important to agree on skill metrics and reference datasets but also query scientific correctness and skill levels in the tails of probability distributions where extreme events lie. It will be important to perform evaluation at user application level where societal impacts are created. 
The effect will be that: 
\begin{itemize}
    \item Model evaluation becomes more versatile and accessible, easier to perform, and include both numerical and ML models.
    \item Easier model evaluation will increase usability for AI agents to develop and test state-of-the-art models.
    \item Model development cycles become shorter, and model development can be coordinated more easily between centres.
    \item Both ML and numerical model evaluation standards benefit from co-developed insights into performance, physical correctness constraints and data statistics, including their limitations.
    \item The risk of overfitting of both ML and physical models will grow as evaluation has a larger impact on model development. Here overfitting refers to models behaving well in all parts that are tested which leads to trust that is not warranted for other parts that remain untested during development.
\end{itemize}

\section{Computing: interactive \& distributed}\label{section:compute}
The key concerns for computing in weather prediction are the size, reliability \& urgency of the compute tasks and the location where the computing is performed. This includes the assurance that computing is available with high reliability (99\%) and that data handling requirements are commensurate with computing. For operational production, interactive and decentralized workflows must still meet hard requirements for 24/7 continuity, controlled latency, monitoring, audit trails, reproducibility, traceability, quality assurance and fallback options. Multi-platform strategies improve resilience and reduce dependency on individual hardware providers.

The fact that compute nodes have become much more powerful over decades is well known, and NWP centres have greatly invested in adapting to new technologies to benefit from what digital technology companies provide -- primarily driven by other markets and, recently, ML \citep{bauer2021digital}. Today, a single GPU node achieves several hundred teraflops of single-precision performance. For trained ML forecast inference, high-profile global weather models have already demonstrated forecasts in seconds to minutes on GPU or TPU hardware \citep{pangu,lam2023learning,price2025probabilistic,neuralgcm}. This does not imply that complete operational production chains move to one node: ensembles, coupled ESMs, data assimilation, and the collective effort to run quality control, I/O, archiving, dissemination, monitoring and fallback still require resilient HPC-scale infrastructure.

Advanced very-high-resolution ensembles of physical models will always require extraordinary HPC capabilities \citep{hoefler2023earth} but remain instrumental for evolving our scientific understanding, developing reference datasets and defining our evaluation framework for ML models. However, the improved availability of Earth-system datasets and the easier access to much larger datasets via data compression and data curation will make it possible and easy to train and initialise ML models at all dates over past decades and along climate model trajectories over the next decades. ML models offer new ways for exploiting distributed HPC capacities to most cost-effectively place both urgent and science advancing tasks. A revised management of location and lifecycle of data and compute is an important ingredient and agentic AI may again play an important role here.

On the user end, improvements in data visualisation will soon allow for a much more detailed study of model output and human interaction \citep{bauer2024digital}. Once data location, pre-processing, data compression and input/output-pipelines are fully optimized, three-dimensional visualisation can be performed locally in real time and interactively including the possibility to support the visualisation via virtual reality. ML can provide context to the scientific information of the data that is currently visualised through the use of LLMs, for example to answer the question whether a change in mean temperature will increase infestations of Bark Beetles in a specific region or whether other pests are likely, given a visualised future weather evolution. 

Fully interactive simulations, i.e. simulations where users can change and rerun with new settings, ideally with a fully integrated visualisation interface, will be much easier to implement with ML than physical models. Initiatives such as Forecast-In-A-Box codeveloped in DestinE already show how easily ML-models can be deployed locally \cprotect\footnote{\verb|https://forecast-in-a-box.ecmwf.int|}.

The effect will be that: 
\begin{itemize}
    \item Progress in digital technologies will contribute substantially to the effectiveness of ML-model development and impact creation.
    \item Distributed computing and data handling concepts help to support both model development and operation, while offering a more cost-effective alternative than today’s insular solutions. 
    \item More decision power and higher productivity is placed near users through interactive, on-the-fly visualisation, verification and interpretation of data, for example, powered by a browser on a single-node on a small cluster or in the cloud.
    \item For many applications data can be recreated (in particular with ML emulators) rather than stored, thus reducing costly data movements.
\end{itemize}

\section{Information extraction: generative methods \& data creation}\label{section:analytics}
Many of the next-generation generative ML models will be able to create missing information as they will learn to replicate the behaviour of model physics rather than predict data. In particular foundation models such as those developed in the WeatherGenerator project\cprotect\footnote{\verb|https://weathergenerator.eu|}, that are trained from self-supervised and masked token learning, will have a very sophisticated latent space that will allow to generate meaningful information from minimal input. First examples of such models are AtmoRep \citep{atmorep}, Aurora \citep{aurora} and the CBottle climate generator \citep{brenowitz2025climate}.

The potential use of these tools goes beyond simple applications such as data gap filling, forecast emulations or ensemble generation. The tools will be able to generate a meaningful Earth-system state from a small amount of meta information and data that constrains the physics such as a date, a sea surface temperature field, or a request such as “I want to see a realistic and strong tropical cyclone that makes landfall in Haiti on the first of January 2030”. As the generative model can produce a set of answers, it is possible to evaluate the probability distributions for specific events, and show how an individual event like this particular tropical storm could play out.

Such models will allow for a totally new way to evaluate climate statistics and extremes. This is because they offer playing through possible scenarios of future events and investigating the leading contributions with much reduced effort. At present, such studies are performed as so-called story lines where either past extreme events are deterministically simulated under future warming scenarios to understand how they would change \citep{athanase2024climate} or how physically plausible global circulation pattern changes would statistically project onto regional impacts \citep{zappa2017storylines}. 

WeatherGenerator-type models will allow such a study interactively on the fly. They will make the biggest difference once combined with interactive visualisation (see Section~\ref{section:compute}). A tool that can generate three-dimensional weather situations for the entire globe will not only be fun to play with, but also extremely useful for teaching--for example when different storm scenarios can be generated by students through perturbations of physical fields without waiting time. 

However, generative machine learning has a very significant catch: As these tools will be trained from a large number of datasets, including observations, reanalyses and free-running model simulations, it will be almost impossible to quantify how reliable a signal is, for example, in the context of a change of certain events with climate change. This will make it very hard to quality-control the results and require more traditional techniques to be maintained in parallel. It will require the weather and climate community to resist the temptation to just accept visualizations as truth and scrutinize the physical realism and quantify uncertainties as in classical approaches. 

The effect will be that:
\begin{itemize}
    \item What-if scenarios, climate statistics and uncertainty propagation can be studied in real time using generative machine learning tools.
    \item ML tools will allow to study massive datasets interactively, helped by ML-supported data compression and visualization.
    \item Testing the physical realism and statistical significance for output from generative models requires a new level of scrutiny to distinguish trustworthiness from hallucination.
\end{itemize}

\section{Outlook}

The topics discussed above have the potential to change how we work and may come with significant changes in infrastructure needs to support model developers, scientists and external users. NWP centres may benefit from adapting proactively as new skills emerge and the cycle of turning innovation into services accelerates and diversifies. Considering these topics together may help organisations navigate this adaptation more effectively.

To maintain transparency, scientific credibility and long-term sustainability, the role of weather and climate centres may broaden rather than fundamentally shift focus: public forecasts, warnings, climate information and service delivery are likely to remain core missions, while centres may increasingly act as providers of trusted datasets, open tools, benchmarks and standards that enable collaboration across research institutions, public services, open communities and companies. Such a broader role may also contribute to national sovereignty, strategic autonomy and global equity as prediction capabilities become increasingly distributed across public centres, private companies and open communities.

In one possible future, reanalysis products, model simulations and observations could increasingly serve as data sources that feed into federated datasets that are more easily accessible and usable. Unified data access formats that support ML and interactions with AI agents, together with globally available metadata describing availability and quality, may become important enablers. ML also has the potential to enhance two-way interactions between end-users and providers of weather predictions and products. Timely and relevant user feedback on forecast performance has considerable potential for reinforcement-learning approaches and service improvement.

NWP centres may increasingly seek to provide capabilities to run both global/local ML and physical models across space and time, together with access to datasets that combine low latency with assured quality. Advanced tools for visualisation and diagnostics, including community-supported benchmarks, could become important components of these capabilities. Achieving this vision may be challenging for individual centres acting alone and could benefit from more distributed approaches to science, coding, tooling, data handling and computing. In particular, data handling and computing may benefit from greater cost-effectiveness and agility at the community level \citep{bauer2024if}. EuroHPC\cprotect\footnote{\verb|https://www.eurohpc-ju.europa.eu/index_en|}, Japan's RIKEN\cprotect\footnote{\verb|https://www.r-ccs.riken.jp/en/|} and the United States Department of Energy\cprotect\footnote{\verb|https://www.energy.gov/science/ascr/advanced-scientific-computing-research|} are likely to remain important resources for extreme-scale simulations that contribute to scientific progress. Continued access to reliable allocations on such systems could support research, the exploration of new hardware technologies at scale and the rapid production of reference and training datasets.

If such developments materialise, users could gain easier access to data, tools and computing resources and apply this infrastructure to their specific applications \citep{koldunov2024local}. They may contribute application-specific data to shared data pools, run both physical and ML models interactively and at scale, and explore uncertainties through advanced visualisation and diagnostic tools. Lower barriers to access could encourage broader use of Earth-system information, including by users with less specialised expertise. In such a future, distinctions between ML-generated and traditionally generated data, models and services may become less pronounced.

However, these developments may also introduce risks, as the traceability of data and information generation could be reduced, quality control may become more complex and opportunities for misuse may increase. Realising the benefits of ML while managing these risks will require continued attention to ethical, governance and quality-assurance frameworks, areas in which the weather and climate community is well positioned to contribute \citep{mcgovern2022we}\cprotect\footnote{\verb|https://wmo.int/media/announcement/wmo-call-artificial-intelligence-and-machine-learning-out|}.

A central risk is the possible erosion of expert knowhow if agents become integral to scientific workflows and operational critical paths. Weather and climate centres may therefore wish to consider mechanisms that help maintain human expertise, including transparent model diagnostics, reproducible experiments, regular exposure of scientists to operational and diagnostic work, interpretable interfaces, adversarial stress tests of model and workflow failures, and practices that preserve the ability to explain why a system works or fails.

While some developments seem easier to anticipate than others, considerable uncertainty remains around the future role of agentic AI and its influence on the working practices of software engineers, scientists and managers. However, some steps to support the safe adoption of agentic AI are more obvious, including standardised data access, the modularisation of model components, and the establishment of automated verification combined with continuous testing. 

The weather and climate community has repeatedly demonstrated its ability to adapt to technological change and is well positioned to shape this transition in ways that create value for society while preserving scientific integrity and operational trust.

\section*{Author contribution}
PD and PB conceptualised this work, and all authors co-wrote this paper. 

\section*{Competing interests}
The authors declare that there are no competing interests. 


\section*{Declaration of generative AI and AI-assisted technologies in the manuscript preparation process}

During the preparation of this work the authors used ChatGPT to make suggestions for improvements of the first draft. After using this tool, the authors reviewed and edited the content as needed and take full responsibility for the content of the published article.

\section*{Acknowledgements}
The authors would like to thank Tim Hunter, Tiago Quintino, Ioan Hadade, Florian Pappenberger, Emma Pidduck and Tony McNally (all ECMWF) as well as Martin Schultz (Juelich Supercomputing Centre) for their helpful discussions and comments. Peter Dueben and Jørn Kristiansen acknowledge funding from the WeatherGenerator Horizon Europe project (grant No 101187947). Peter Dueben acknowledges funding from the ESiWACE3 EuroHPC-JU project (grant No 101093054).

\bibliographystyle{plainnat}

\bibliography{JEMS_2026}

\end{document}